\documentclass[3p,times]{elsarticle}

\usepackage{ecrc}

\volume{00}

\firstpage{1}

\journalname{Nuclear Physics A}

\runauth{}

\jid{nupha}

\usepackage{amssymb}

\usepackage[figuresright]{rotating}

\newcommand{\lsim}{\,{\buildrel < \over {_\sim}}\,}
\newcommand{\gsim}{\,{\buildrel > \over {_\sim}}\,}
\newcommand{\sqrtsNN}{\sqrt{s_{\scriptscriptstyle{{\rm NN}}}}}

\newcommand{\gev}{\mathrm{GeV}}
\newcommand{\tev}{\mathrm{TeV}}
\newcommand{\fm}{\mathrm{fm}}

\newcommand{\cm}{\mathrm{cm}}

\newcommand{\mum}{\mathrm{\mu m}}

\newcommand{\PbPb}{\mbox{Pb--Pb}}

\newcommand{\pt}{p_{\rm t}}
\renewcommand{\d}{{\rm d}}
\newcommand{\dEdx}{{\rm d}E/{\rm d}x}

\newcommand{\Jpsi} {\mbox{J\kern-0.05em /\kern-0.05em$\psi$}\xspace}

\begin{document}

\begin{frontmatter}



\dochead{}

\title{Measurement of heavy-flavour production in proton--proton collisions at $\sqrt{s}=7~\tev$ with ALICE}


\author{A. Dainese}

\address{INFN -- Sezione di Padova, via Marzolo 8, I-35131, Padova, Italy}

\begin{abstract}
  The measurement of the heavy-flavour production cross sections in pp collisions
at the LHC will allow to test perturbative QCD calculations in a new
energy domain. Moreover, within the physics program of the ALICE experiment, 
it will provide the reference for the study 
of medium effects in Pb--Pb collisions, where heavy quarks are regarded
as sensitive probes of parton--medium interaction dynamics.
We present the status and first preliminary results of charm and beauty
    production measurements with the ALICE experiment, using hadronic D meson
  decays and semi-leptonic D and B meson decays, including the first cross section measurement
  of muons from heavy flavour decays at forward rapidity.  
We also describe the preliminary cross section measurement for $\rm J/\psi$ 
production, obtained using the
di-electron decay channel at central rapidity and the di-muon decay channel at forward
rapidity.
\end{abstract}


\end{frontmatter}


\section{Introduction}
\label{sec:intro}

ALICE~\cite{alicePPR1,alicePPR2} is the dedicated heavy-ion experiment at the Large
Hadron Collider (LHC). The main physics goal of the experiment is the 
study of strongly-interacting matter in the conditions of high energy 
density ($>10~\gev/\fm^3$) and high temperature ($\gsim 0.3~\gev$),
expected to be reached in central \mbox{Pb--Pb} collisions at a centre of mass 
energy $\sqrtsNN$ of a few TeV per nucleon--nucleon pair.
Under these 
conditions, according to lattice QCD calculations, quark confinement into 
colourless hadrons should be removed and
a deconfined Quark--Gluon Plasma should be formed~\cite{karsch}.
The ALICE experiment has started taking data with proton--proton collisions at the LHC
at $\sqrt{s}=0.9~\tev$ in November 2009 and at 7~TeV in March 2010. The 
first Pb--Pb run at the LHC, with $\sqrtsNN=2.76~\tev$, took place successfully from 
November 7 to December 6, 2010.

Heavy-flavour particles
are regarded as effective probes of the conditions of the system produced in nucleus--nucleus collisions. In particular: 
\begin{itemize}
\item open charm and beauty hadrons would be sensitive to the energy density,
through the mechanism of in-medium energy loss of heavy quarks;
\item quarkonium states would be sensitive to the initial temperature of the
system, through their dissociation due to colour screening;
\item initially uncorrelated charm and anti-charm quarks, 
abundantly produced in the initial stage of the collision,
may recombine and yield an increase in the number of observed charmonium particles.
\end{itemize}
More details on the physics motivation for these measurements in Pb--Pb collisions
can be found in Ref.~\cite{alicePPR1,alicePPR2}. In this report, we describe the first results
on open and hidden (quarkonium) heavy flavour production in pp collisions at $\sqrt{s}=7~\tev$.

Heavy-quark production measurements in proton--proton collisions 
at LHC energies, 
besides providing the necessary baseline for
the study of medium effects in nucleus--nucleus collisions, are interesting 
{\it per se}, as a test of perturbative QCD in a new energy domain. State-of-the-art implementations of perturbative QCD calculations describe well the beauty production cross
section measured in $\rm p\overline p$ collisions at $\sqrt{s}=1.96~\tev$ at the 
Tevatron~\cite{fonllBcdf}. Also the production of charm hadrons (D mesons) is reproduced within
the theoretical uncertainties of the calculation~\cite{charmcdf}. However, in this case the comparison suggests that charm production is slightly underestimated in the calculations,
as observed also in pp collisions at RHIC at the much lower $\sqrt{s}=0.2~\tev$~\cite{phenixelepp,starelepp}.
Moving from open to hidden heavy flavour production, the hadroproduction of quarkonium states is a process where QCD is involved in both perturbative and non-perturbative aspects. Various models~\cite{brambilla} have been proposed to describe the results obtained at the Tevatron~\cite{cdfjpsi,d0jpsi}, but they fail to reproduce simultaneously the production cross sections, the transverse momentum distributions, and the measured polarization, as well as their dependence on rapidity.

In the next section, we describe the ALICE heavy flavour measurements program. The strengths of ALICE, also in comparison with the other LHC experiments, are: (i) the large acceptance at 
low transverse momentum $\pt$ (starting from 0 for $\rm J/\psi$ measurements and  2--$3~\gev/c$
for open flavour) and (ii) in rapidity, covering both central and forward regions; (iii) the capability 
to measure various final states (hadronic, electronic, and muonic).  
The aspects of 
the experimental apparatus that are more closely related to these measurements are also described in the next section, along with the
data taking conditions for pp in 2010. Then, we give the status and results on D meson reconstruction (section~\ref{sec:Dmesons}), single-electron and single-muon measurements
(sections~\ref{sec:ele} and~\ref{sec:muon}), and $\rm J/\psi$ production (section~\ref{sec:jpsi}).

\section{Heavy-flavour detection in the ALICE apparatus and pp data taking in 2010}
\label{sec:exp}

The ALICE experimental setup, described in detail in~\cite{aliceJINST},
allows the detection of open charm and beauty hadrons and of quarkonia
in the high-multiplicity environment 
of central \PbPb~collisions at LHC energy, in which a charged particles rapidity density
$\d N_{rm ch}/\d\eta\approx 1600$ was measured at $\sqrtsNN=2.76~\tev$~\cite{multPbPb}. 
The heavy-flavour capability of the ALICE detector is mainly provided by:
\begin{itemize}
\item Tracking system; the Inner Tracking System (ITS), 
the Time Projection Chamber (TPC), and the Transition Radiation Detector (TRD),
embedded in a magnetic field of $0.5$~T, allow track reconstruction in 
the pseudorapidity range $-0.9<\eta<0.9$ 
with a momentum resolution better than
2\% for $\pt<20~\gev/c$ 
and a transverse impact parameter\footnote{The transverse impact parameter,
$d_0$, is defined as the distance of closest approach of the track to the 
interaction vertex, in the plane transverse to the beam direction.} 
resolution better than 
$75~\mum$ for $\pt>1~\gev/c$ 
(the two innermost layers of the ITS, $r\approx 3.9$ and $7.6~\cm$, 
are equipped with silicon pixel 
detectors).
\item Particle identification system; charged hadrons are separated via 
their specific energy deposit $\dEdx$ in the TPC and via time-of-flight measurement in the 
Time Of Flight (TOF) detector; 
electrons are identified at low $\pt$ ($<4~\gev/c$) via the $\dEdx$ in the TPC and the 
time of flight in the TOF, at intermediate $\pt$ (1--$10~\gev/c$) in the dedicated Transition Radiation Detector (TRD) and at large $\pt$ ($>5$--$10~\gev/c$) 
in the Electromagnetic Calorimeter (EMCAL); 
muons are identified in the muon 
spectrometer covering the pseudo-rapidity range $-4<\eta<-2.5$. 
\end{itemize}

The main analyses, ongoing with pp data and in preparation for the Pb--Pb case, are:
\begin{itemize}
\item Open charm (section~\ref{sec:Dmesons}): fully reconstructed hadronic decays 
$\rm D^0 \to K^-\pi^+$, $\rm D^0 \to K^-\pi^+\pi^-\pi^+$, 
$\rm D^+ \to K^-\pi^+\pi^+$, $\rm D^{*+}\to D^0\pi^+$,
$\rm D_s^+ \to K^-K^+\pi^+$, $\rm \Lambda_c^+ \to p K^-\pi^+$ in $|\eta|<0.9$.
In this report we present the $\pt$ distributions of the $\rm D^0$ and $\rm D^+$ mesons
in $3<\pt<12~\gev/c$.
\item Open charm and beauty (sections~\ref{sec:ele} and~\ref{sec:muon}): 
inclusive single leptons ${\rm D,\,B\to e}+X$ 
in $|\eta|<0.9$ and ${\rm D,\,B\to\mu}+X$ in $-4<\eta<-2.5$; inclusive displaced
charmonia ${\rm B\to J/\psi\,(\to e^+e^-)}+X$ in $|\eta|<0.9$. In this report we present
the $\pt$-differential cross section of inclusive muons from heavy flavour decays in $2<\pt<6.5~\gev/c$. 
\item Quarkonia (section~\ref{sec:jpsi}): $\rm c\overline c$ (J/$\psi$, 
$\psi^\prime$) and $\rm b\overline b$ ($\Upsilon$, 
$\Upsilon^\prime$, $\Upsilon^{\prime\prime}$) states 
in the ${\rm e^+e^-}$ ($|\eta|<0.9$) and $\mu^+\mu^-$ ($-4<\eta<-2.5$) 
channels. In this report we present
the rapidity-differential cross section for $\pt>0$ of inclusive $\rm J/\psi$ and 
the $\pt$-differential cross section at forward rapidity in $0<\pt<8~\gev/c$.
\end{itemize} 

The results that we present are obtained from data recorded during the first 2--3 months of 
the pp run at $\sqrt{s}=7~\tev$. The main trigger selection was a minimum-bias one, defined by the presence of at least a signal in either of two scintillator hodoscopes, the V0 detectors, 
positioned in the forward and backward regions of the experiment, or in the silicon pixel barrel
detector. This trigger selects events with at least one charged particle in eight pseudorapidity 
units and is sensitive to about 95\% of the pp inelastic cross section. A second trigger selection
requests, in addition to the minimum-bias condition, 
at least one muon trigger signal ($\pt>0.5~\gev/c$). The results presented correspond to 
about $10^8$ minimum-bias triggers (integrated luminosity: $1.4~\rm nb^{-1}$) and 
$10^7$ muon triggers ($11.6~\rm nb^{-1}$).

\section{$\rm D^0$ and $\rm D^+$ meson $\pt$ distributions $\d N/\d\pt$}
\label{sec:Dmesons}

\begin{figure}[!t]
  \begin{center}
  \includegraphics[width=0.49\textwidth]{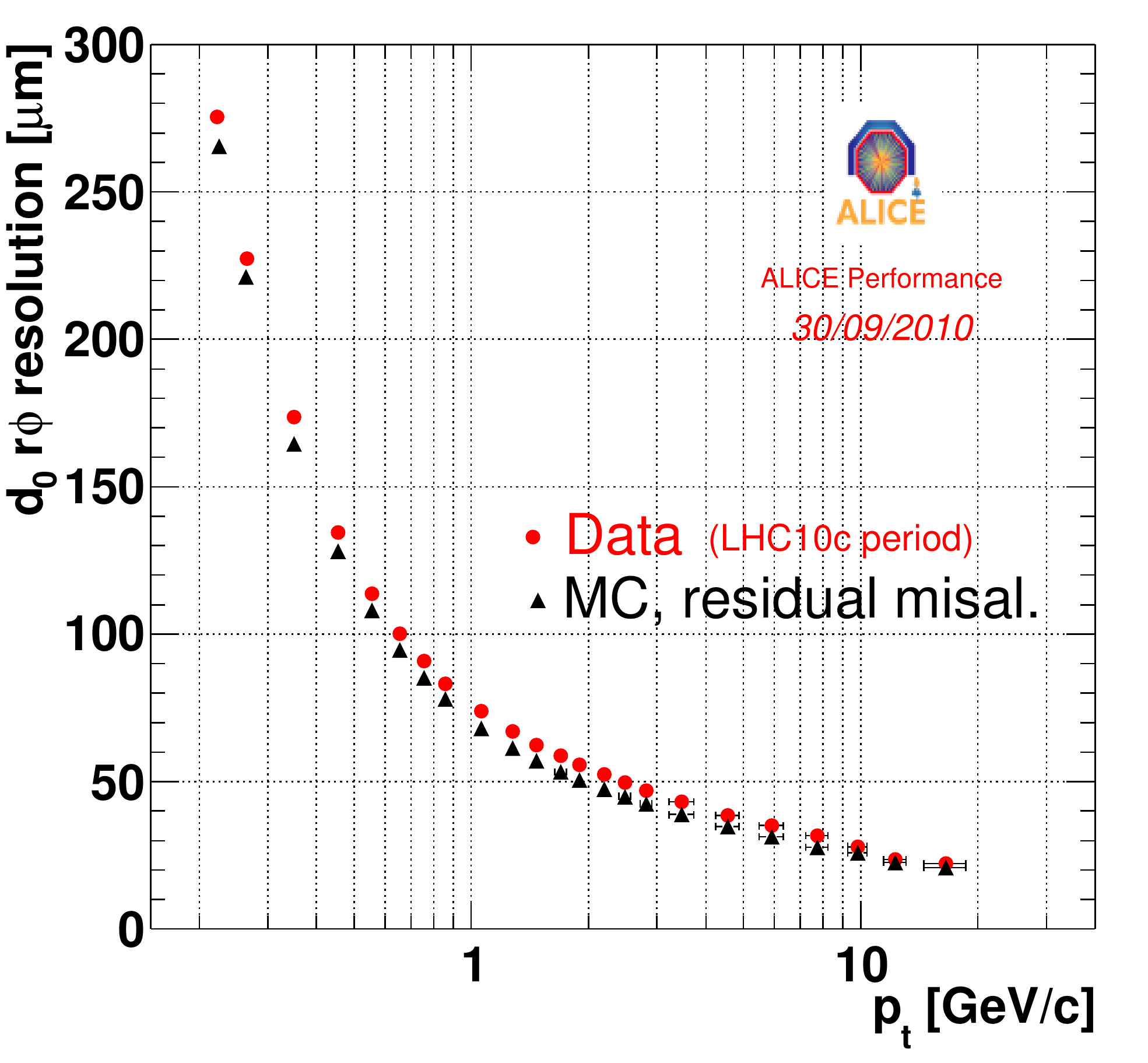}
  \includegraphics[width=0.49\textwidth]{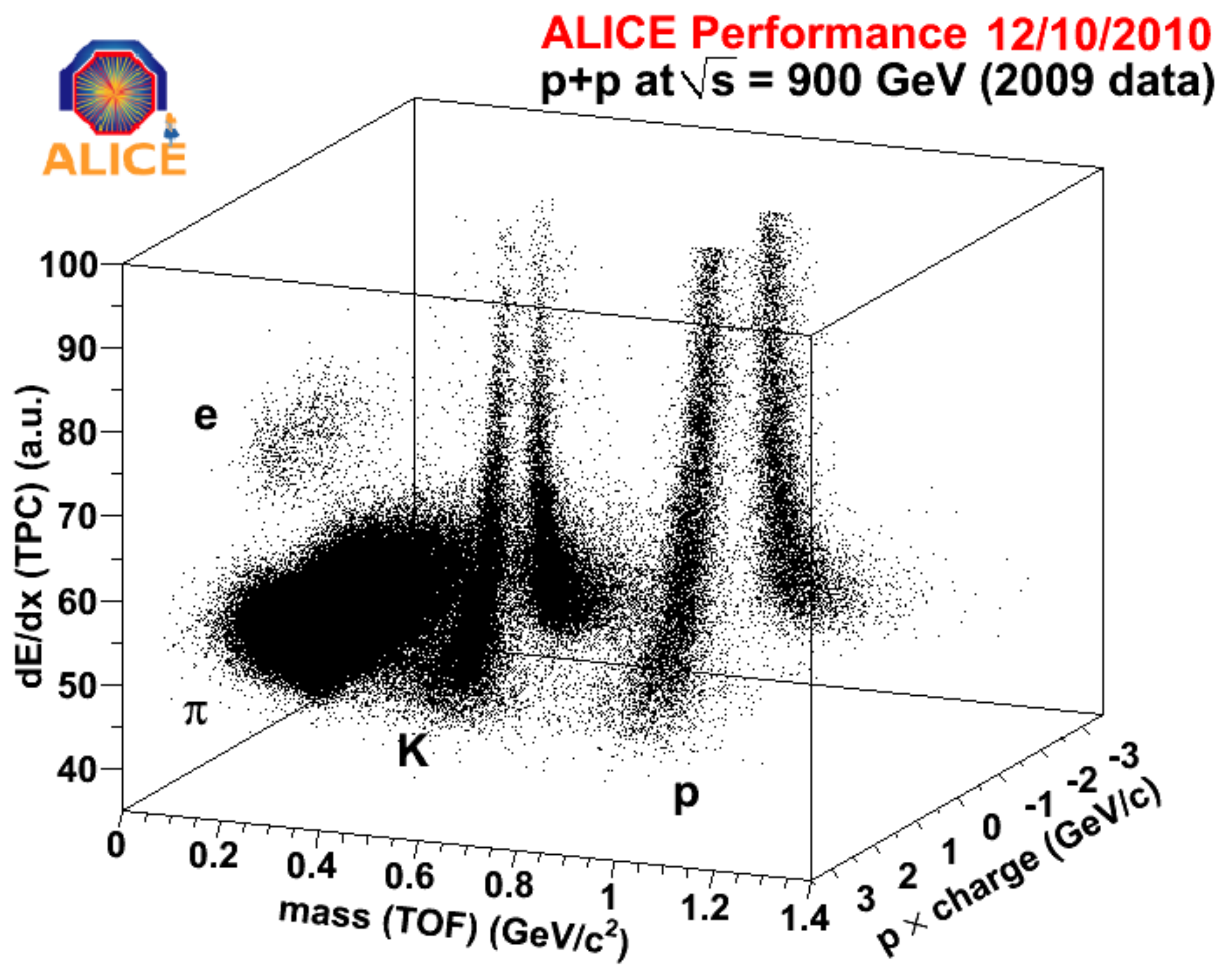}
  \caption{ALICE detector performance results relevant for heavy-flavour measurements.
              Left: the track impact parameter resolution in the transverse plane ($r\phi$ direction)
              as a function of $\pt$; the results for data and simulation are compared; in both cases the resolution is obtained as the sigma of a gaussian fit to the central part of the inclusive impact parameter distribution of tracks reconstructed with at least 70 associated hits in the TPC and
              2 associated hits in the pixel detector; this sigma includes the resolution on primary vertex position, which is reconstructed without the track under test, to avoid a bias.
Right: correlation of the PID signals in the TPC ($\dEdx$) and TOF (time of flight) detectors, as a 
function of track momentum.}
\label{fig:performance}
\end{center}
\end{figure}

\begin{figure}[!t]
  \begin{center}
  \includegraphics[width=0.44\textwidth]{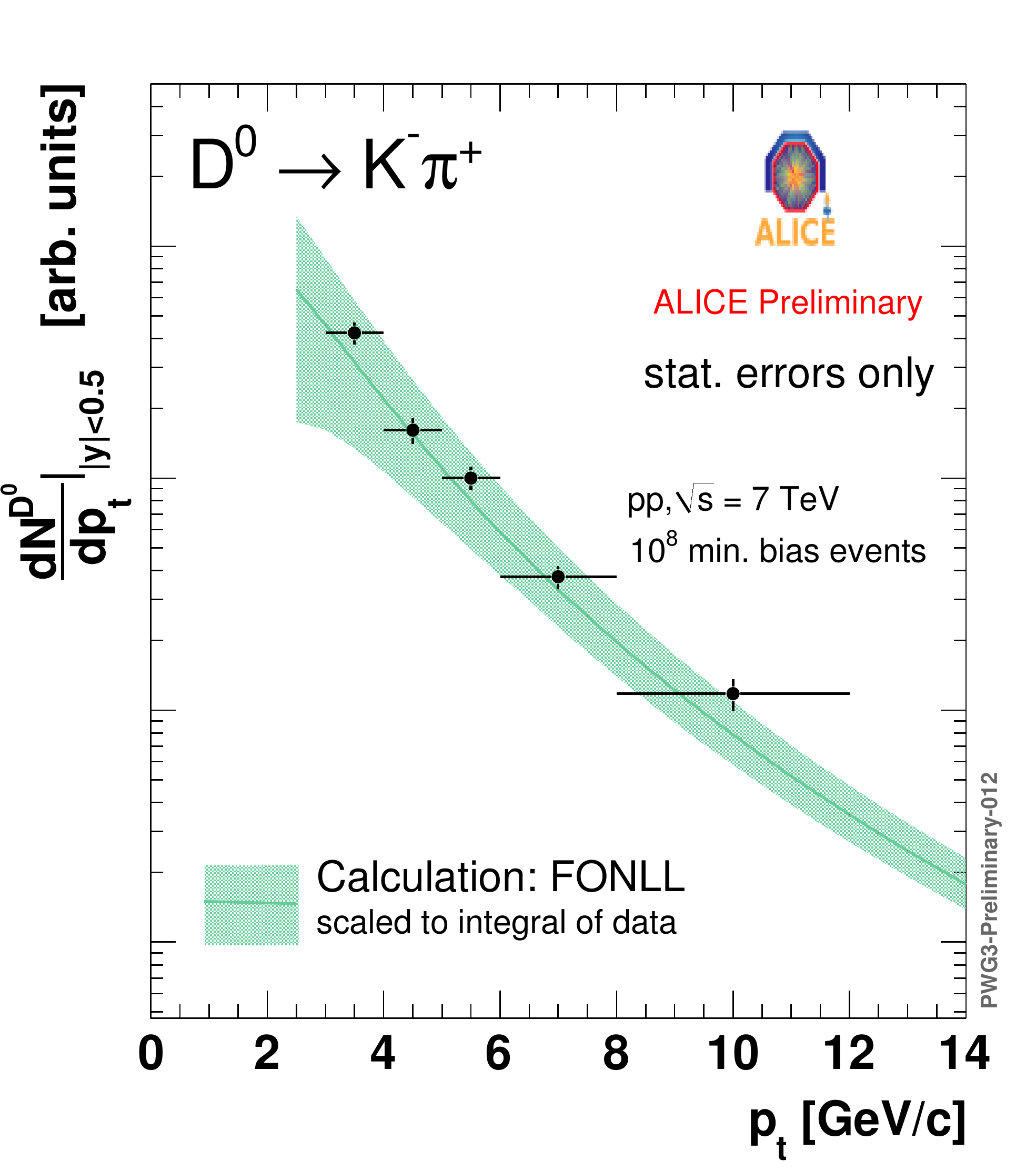}
  \hfill
  \includegraphics[width=0.44\textwidth]{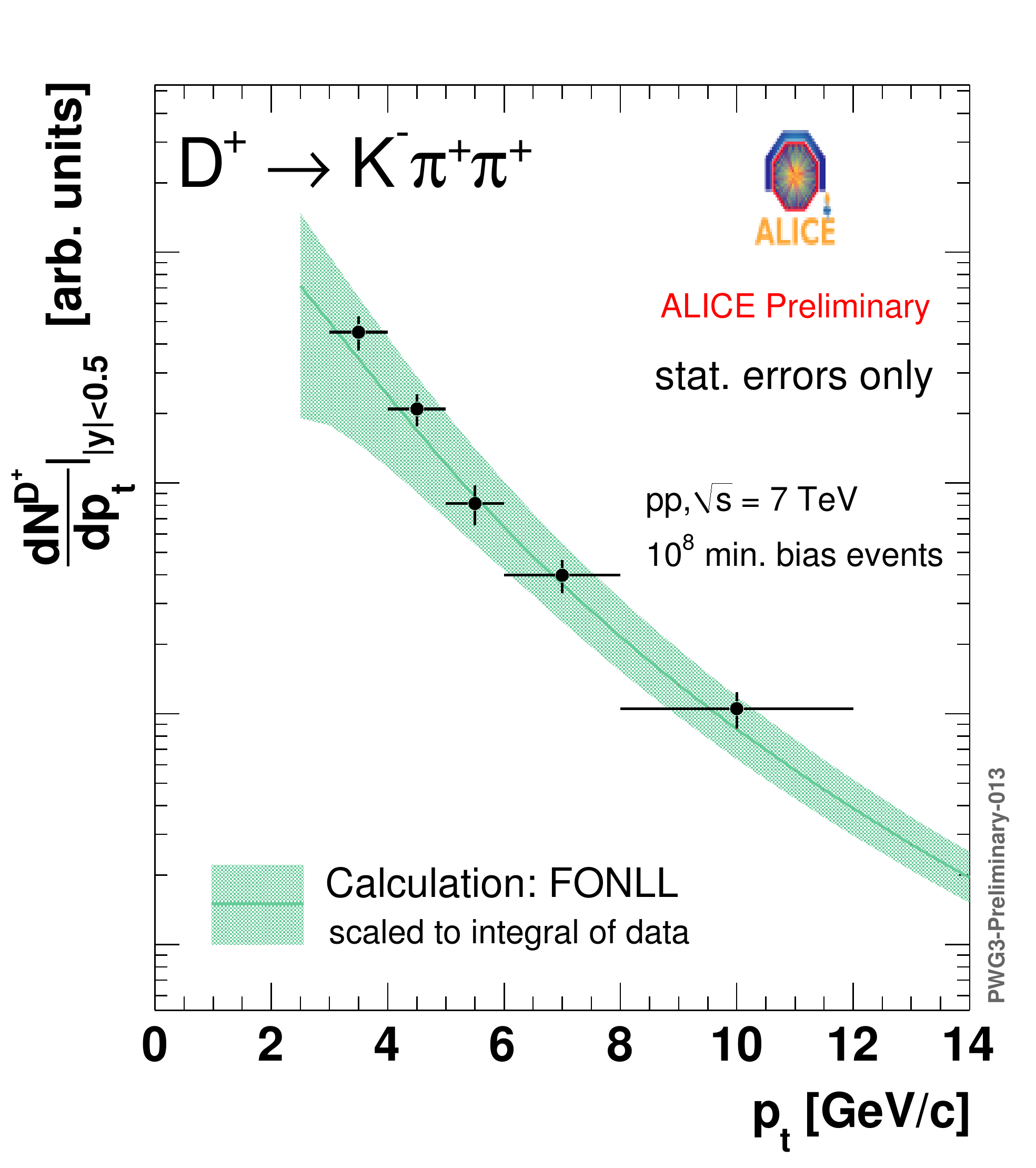}
  \caption{Transverse momentum distributions of $\rm D^0$ and $\rm D^+$ mesons, shown with statistical errors only and arbitrary normalization, and compared to the shape predicted by the FONLL pQCD calculation~\cite{fonllpriv}.}
\label{fig:DdNdpt}
\end{center}
\end{figure}

Among the most promising channels for open charm detection are the 
$\rm D^0 \to K^-\pi^+$ ($c\tau\approx 120~\mum$, branching ratio 
$\approx 3.8\%$) and $\rm D^+ \to K^-\pi^+\pi^+$ ($c\tau\approx 300~\mum$, 
branching ratio $\approx 9.2\%$) decays. The detection strategy 
is based on the selection of displaced-vertex topologies, i.e. the separation of tracks
from the secondary vertex from those originating in the primary vertex, 
and good alignment between the reconstructed D meson momentum 
and flight-line. 
The secondary vertex is reconstructed using good-quality tracks with $|\eta|<0.8$ and at least 70, out of a maximum of 160, associated hits in the TPC and at least one associated hit in the pixel detector (this selection leads to an average of 4.7 associated hits in the 6 ITS layers). 
The displaced-secondary-vertex selection 
strategy relies on the excellent tracking precision provided, in particular, by the 
ALICE Inner Tracking System. This silicon tracker was aligned using survey measurements and 
tracks from cosmic-ray particles and pp collisions, based on the methods described in 
Ref.~\cite{itsalignment}, and the resolution achieved on the track impact parameter measurement,
shown as a function of $\pt$ in Fig.~\ref{fig:performance} (left), is close to the design goal, with
values of $75\,(20)~\mum$ at $1\,(20)~\gev/c$, including the primary vertex position resolution.
The efficiency of the secondary vertex selection ranges from 1--2\% to 15\% in the range $2<\pt<12~\gev/c$.
The identification of the charged kaon in the TPC and TOF detectors helps to further reduce the
background in the low transverse momentum region. The separation power using the combined response from the two detectors is qualitatively illustrated in Fig.~\ref{fig:performance} (right).
In order to ensure a negligible systematic error from the PID response, we apply a strategy that grants an efficiency close to 100\% for the signal, i.e. we reject a candidate D meson if
all decay tracks do have a PID signal associated to them and all these signals are incompatible 
with the expected signal from a charged kaon.
An invariant-mass analysis is used to extract the raw signal 
yield, to be then corrected for detector acceptance and 
for PID, selection and reconstruction efficiency, evaluated from a detailed detector simulation. 
The contamination of D mesons from B meson
decays is currently estimated to be of about 15\%, using the beauty production cross
section predicted by the FONLL (fixed-order next-to-leading log) 
calculation~\cite{fonllpriv} and the detector simulation, 
and it is subtracted from the measured raw $\pt$ spectrum, before applying the efficiency 
corrections (that are calculated for prompt D mesons).

Using a sample of $10^8$ minimum-bias collisions, we have produced the $\pt$ distributions of 
$\rm D^0$ and $\rm D^+$ mesons in the range $3<\pt<12~\gev/c$, which are shown in 
Fig.~\ref{fig:DdNdpt} in arbitrary units, with statistical errors only. The acceptance in rapidity 
for D mesons depends on $\pt$, namely it increases from $|y|<0.5$ at $\pt=0$ to $|y|<0.8$
for $\pt\gsim 5~\gev/c$. For each $\pt$ bin we scale the corrected yield from the average
$y$ acceptance of that bin to a common acceptance $|y|<0.5$, assuming a flat rapidity 
distribution (a valid assumption in the relevant range, $|y|<0.8$).
The measured $\pt$ distributions are compared to the 
shape of the FONLL theoretical predictions~\cite{fonllpriv}. We find that the $\pt$ shapes
are compatible with the pQCD predictions. Using the yields integrated for $\pt>3~\gev/c$,
we calculate a $\rm D^0/D^+$ ratio of $2.5 \pm 0.3\,(stat.)$, in agreement with 
previous measurements, as shown in the left-hand panel of Fig.~\ref{fig:ele}.

The evaluation of the systematic uncertainties, as well as the absolute 
normalization, 
are ongoing. Using the entire data sample collected in 2010, we expect to increase the $\pt$ 
coverage (e.g. the $\rm D^0\to K^-\pi^+$ signal is already observed below $1~\gev/c$) 
and to perform the correction for feed-down from beauty decays using the measured displacement
of feed-down D mesons from the primary vertex. 

The $\rm D^{*+}\to D^0\pi^+$ analysis is also well advanced, with the signal extracted in several 
$\pt$ bins between 2 and $16~\gev/c$. Promising signals 
have also been observed for the channels
$\rm D^0 \to K^-\pi^+\pi^-\pi^+$, $\rm D_s^+ \to K^-K^+\pi^+$, 
and $\rm \Lambda_c^+ \to p K^-\pi^+$. The latter, for which the displaced vertex selection 
is less efficient (the $\rm \Lambda_c$ has a mean proper decay length of only $59~\mum$),
benefits from the excellent proton identification capabilities of the TPC and TOF detectors
in the range $p\lsim 3~\gev/c$.

\section{Electrons from heavy-flavour decays}
\label{sec:ele}

\begin{figure}[!t]
  \begin{center}
  \includegraphics[width=0.44\textwidth]{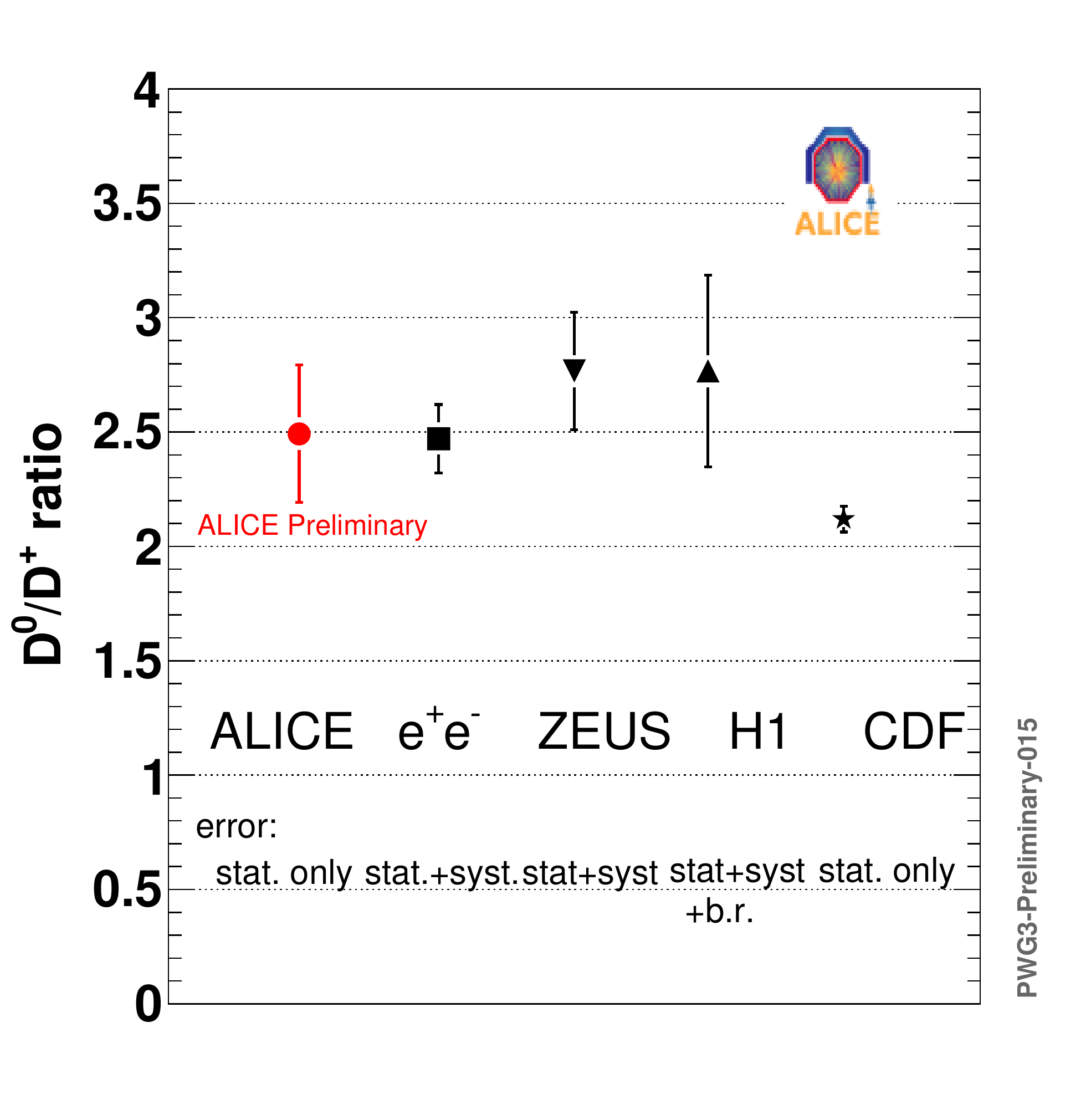}
  \hfill
  \includegraphics[width=0.48\textwidth]{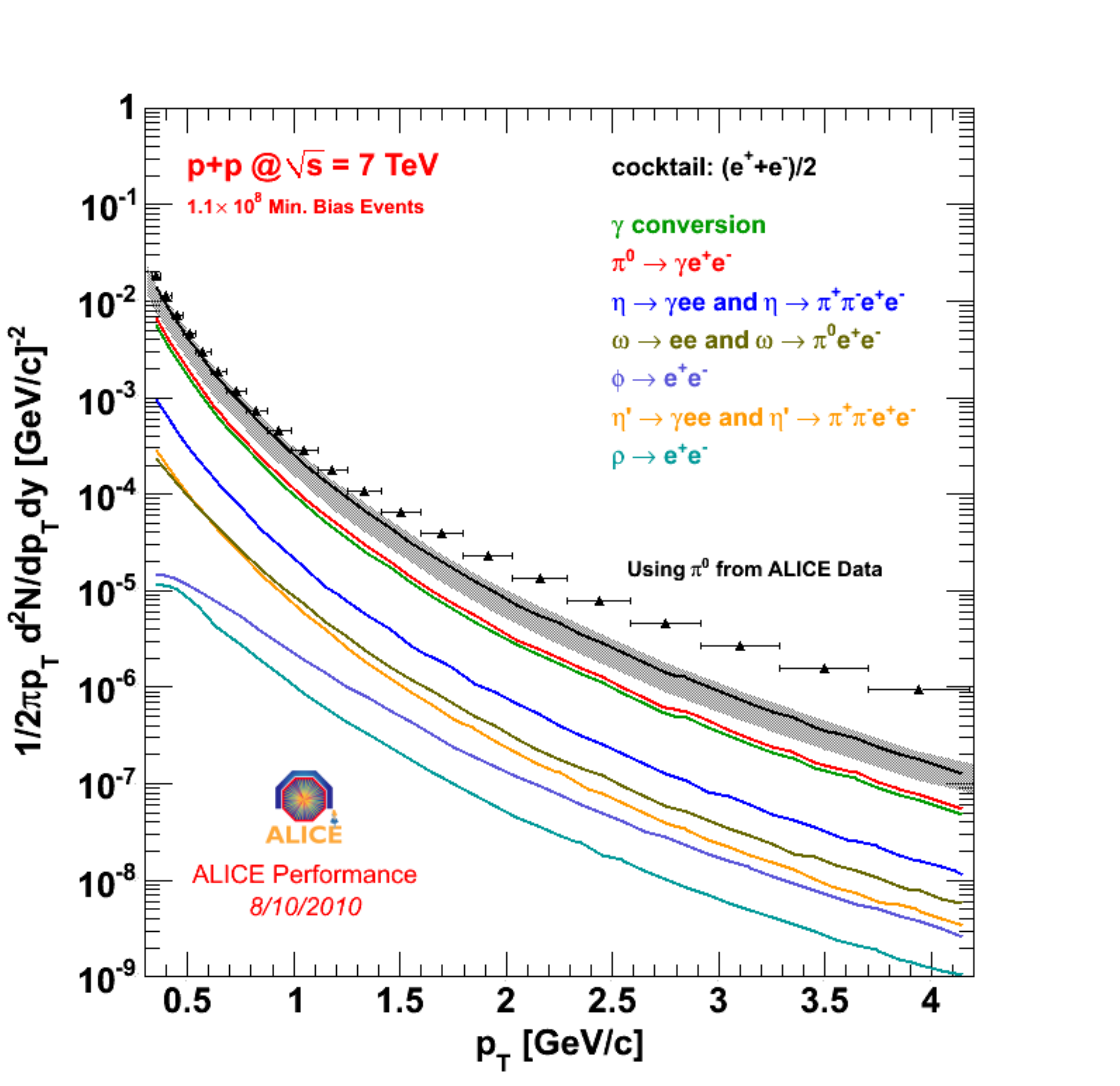}
  \caption{Left: $D^0/D^+$ ratio for $\pt>3~\gev/c$, compared to previous measurements;
               the ratios for the other experiments have been calculated using the data in 
               Refs.~\cite{zeusD,charmcdf}.
               Right: invariant inclusive electron spectrum $\d N/\d\pt$, measured with a sample of 
                $10^8$ minimum-bias pp collisions, and compared to the cocktail of background 
                electrons sources, based on the 
                measured $\pi^0$ cross section~\cite{pi0}.}
\label{fig:ele}
\end{center}
\end{figure}

The production of open charm and beauty can be studied by detecting the 
semi-electronic decays of D and B mesons~\cite{silvia}. 
Such decays have a branching ratio of $\simeq 10\%$ 
(plus 10\% from cascade decays ${\rm b\to c} \to e$, that only populate 
the low-$\pt$ region in the electron spectrum).
The ALICE apparatus
provides excellent electron identification using signals from four of its detectors
at central rapidity. A measurement of the production cross section of heavy-flavour
decay electrons can then be obtained using two methods. 

In the first method a cocktail of background electrons is subtracted from
    the inclusive electron spectrum. The background electrons originate from
    various sources, namely, electrons from light hadron decays (mainly
    $\pi^0$ Dalitz decays, in addition to $\eta$, $\rho$, $\omega$, and $\phi$ decays), 
    photon
    conversions in the material of the beam pipe and of the inner pixel layer,
    direct radiation, and electrons from decays of heavy quarkonia. For $\pt$
    up to a few GeV/$c$ this cocktail can be determined accurately on the basis
    of the measured $\pi^0$ cross section $\d\sigma/\d\pt$. After the cocktail
    subtraction, the inclusive cross section of electrons from charm and
    beauty decays is obtained.

The second method exploits the relatively large decay length of B mesons ($c\tau\approx 500~\mum$): electrons from beauty decay have are typically well-separated from the primary vertex
and a cut on the minimum transverse impact parameter is expected to reject most of the background electrons from the aforementioned sources and a large fraction of the charm 
decay electrons, thus yielding a sample dominated by beauty decay electrons. 

The basis of this measurement is, clearly, a robust electron identification. The initial analysis
focuses on using the TPC and TOF detectors for this purpose, while the TRD and EMCAL 
detectors are being commissioned and calibrated (due to the tight installation schedule,
they could not be commissioned with cosmic-ray tracks, prior to the beginning of the LHC
proton--proton program). As illustrated in the correlation distribution shown in 
Fig.~\ref{fig:performance} (right), electron tracks can be separated from hadron tracks in the momentum range up to a few $\gev/c$. The TOF signal allows to reject most of the kaons and protons, that cross the electron $\dEdx$ band in the TPC at $p\sim 1$--$2~\gev/c$. Then,
a cut on the TPC $\dEdx$ allows to separate the electrons from the pions up to about 
$4~\gev/c$. The residual pion contamination is lower than $15\%$ in this momentum range, 
as measured from the data by fitting with 
a two component function the TPC $\dEdx$ distribution in narrow momentum slices.
Figure~\ref{fig:ele} (right) shows the resulting inclusive electron spectrum in the range 
$0.4<\pt<4~\gev/c$, corrected for acceptance and efficiency, and unfolded to account for the
$\pt$ smearing induced by the Bremsstrahlung effect. The cocktail is also shown. 
This is obtained adopting
as an input the $\pi^0$ cross section measured in ALICE via double-conversion 
reconstruction~\cite{pi0}, assuming $m_{\rm t}$ (transverse mass) scaling to deduce
the other light meson spectra, and using an event generator for the particle decays to electrons. 
Data and cocktail are normalized to one minimum-bias
pp event and are thus comparable. The excess of electrons above the cocktail level can be attributed
to the signal of electrons from heavy-flavour decays (including $\rm J/\psi$, not yet included in
the cocktail). This excess reaches a factor 2--4 for $\pt>2~\gev/c$.
The next steps in this analysis are the normalization of the electron spectrum to the cross
section level (which will allow a first comparison with inclusive heavy-flavour electrons 
predicted by pQCD calculations), the extension to larger $\pt$ using the signals from TRD
and EMCAL, and the selection of displaced electrons (which will allow to obtain
a cross section of beauty-decay electrons).

\section{Cross section of forward single muons from heavy-flavour decays}
\label{sec:muon}

\begin{figure}[!t]
  \begin{center}
  \includegraphics[width=0.58\textwidth]{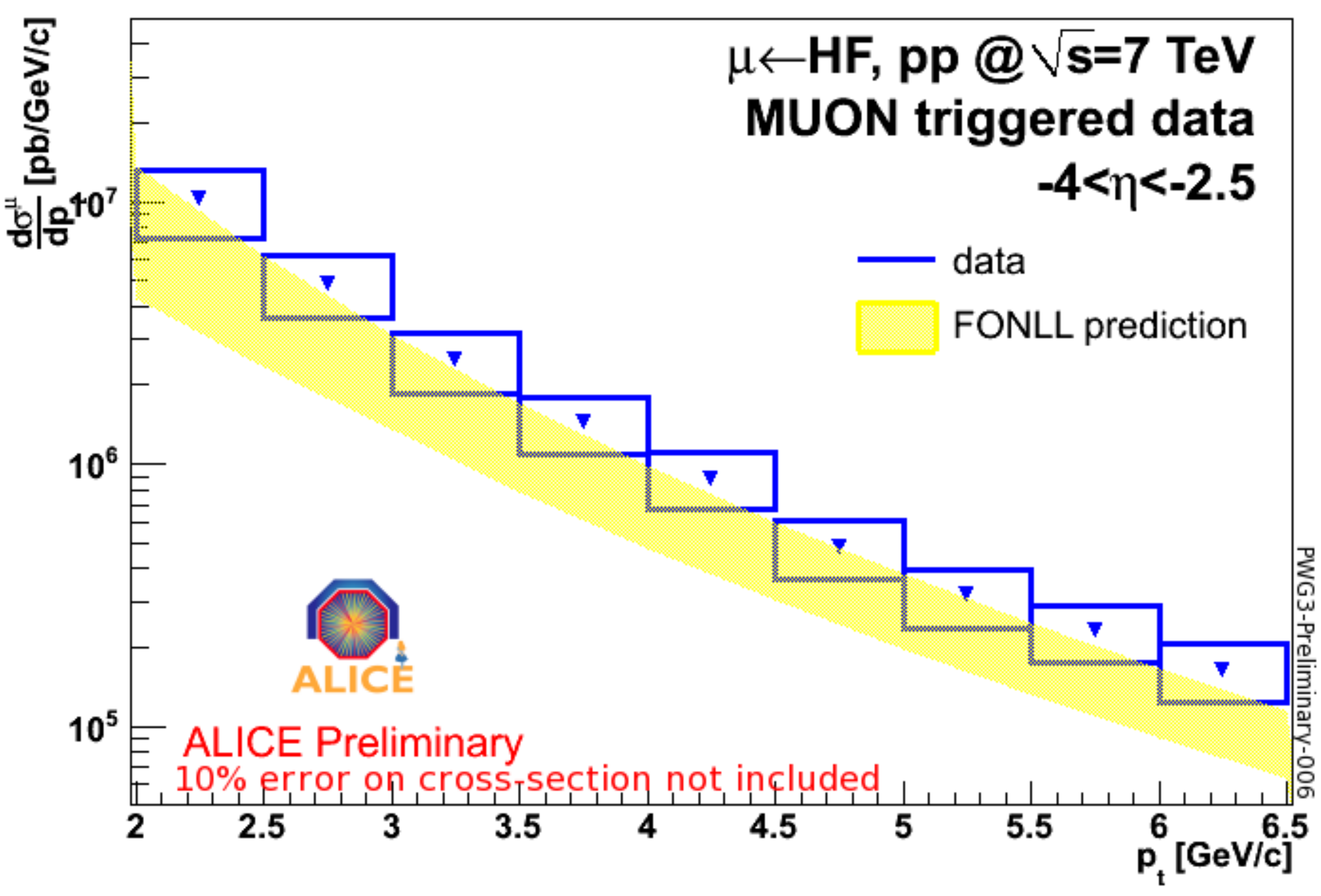}
  \caption{Differential transverse momentum cross section for muons from heavy-flavour decays in $-4 <\eta<-2.5$. The statistical error is small (hidden by the markers). The systematic errors (boxes) do not include an additional 10\% error on the minimum-bias pp cross section. 
  The comparison with the FONLL prediction~\cite{fonllpriv} is also shown.}
\label{fig:muon}
\end{center}
\end{figure}

Heavy-flavour production at forward rapidity can be studied using the single-muon $\pt$ 
distribution~\cite{diego}, measured in the ALICE muon 
spectrometer. Comprising five tracking stations, two trigger stations, a thick frontal 
hadron absorber,
a muon filter iron wall, and a large dipole magnet, 
the spectrometer covers the acceptance $-4<\eta<-2.5$.

The extraction of the heavy-flavour contribution from the single muon spectra requires the subtraction of three main sources of background: a) muons from the decay-in-flight of light hadrons (decay muons); b) muons from the decay of hadrons produced in the interaction with the front absorber (secondary muons); c) punch-through hadrons.
The last contribution can be efficiently rejected by requiring the matching of the reconstructed tracks with the tracks in the trigger system. Due to the lower mass of the parent particles, the background muons have a softer transverse momentum than the heavy-flavour muons, and dominate the low-$\pt$ region. The analysis focuses on the region $2<\pt < 6.5~\gev/c$, 
the upper limit being determined by the $\pt$ resolution of spectrometer with the initial, only partial, 
alignment.  
Simulation studies indicate that, 
in this momentum range, the contribution of secondary muons is small (about 3\%).
In this region, the main source of background consists of decay muons (about 25\%), which have been subtracted by means of Monte Carlo simulations. Namely, 
the transverse momentum distribution from the Perugia-0 tune of PYTHIA~\cite{perugia}
and a detailed simulation of the detector was normalized to the data, in the range $0.5<\pt<1~\gev/c$ where this contribution is dominant according to the simulation and it was subtracted
from the data.
The systematic error introduced by this procedure was estimated as 30\% to 20\% (from low to 
high $\pt$) by comparing the results  obtained with other tunings of PYTHIA and by varying
by 100\% the contribution of secondary particles produced in the frontal and 
beam-pipe absorbers.

After background subtraction, the muon $\pt$ spectrum is corrected for efficiency (which is
larger than 87\%) and normalized to a cross section using the minimum-bias pp cross section
measured, with a "van der Meer scan", as $\sigma_{\rm MB} = 71.4\pm 7.1\,(syst.)~\rm mb$.
The charm and beauty decay muon cross section $\d\sigma/\d\pt$ in the range 
$2<\pt<6.5~\gev/c$ and $-4<\eta<-2.5$ is presented in Fig.~\ref{fig:muon}. The corresponding 
FONLL~\cite{fonllpriv} pQCD
calculation agrees with our data within uncertainties. The next step in this analysis is the 
extension of the high-$\pt$ reach up to about $20~\gev/c$, using data collected since summer
2010 that have been reconstructed with improved alignment corrections for the tracking chambers. The inclusive single-muon cross section is expected to be dominated by beauty 
decays in the range $10<\pt<20~\gev/c$; therefore, this measurement will constitute 
the reference for the study of b quark quenching at forward rapidity in Pb--Pb collisions.

\section{Cross section for inclusive $\rm J/\psi$ production}
\label{sec:jpsi}

$\rm J/\psi$ production can be measured in the ALICE detector 
at central rapidity, using the di-electron decay channel, and at forward rapidity, 
using the di-muon decay channel. In both cases with acceptance down to $\pt=0$.
These measurements are described in detail in Ref.~\cite{roberta}.

For the central rapidity case, for the first analysis with limited statistics (about $10^8$ 
minimum-bias events, i.e. $\rm 1.4~nb^{-1}$) presented in this report, electron identification was based only on 
the TPC $\dEdx$ technique, with a $\pm 3\,\sigma$ inclusion band around the electron
Bethe-Bloch line and two $\pm 3\,\sigma$ exclusion bands for pions and protons. 
Besides the TPC PID cuts, the tracks are required to have a minimum $\pt$ of $1~\gev/c$, 
a minimum of 90 TPC hits, and a hit in the innermost pixel layer (in order to reject electrons from photon conversions). 
The signal is extracted by subtracting from the invariant mass distribution of  
electron pairs with opposite-sign charges the corresponding like-sign background distribution. 
A signal of $95\pm18(stat.)$ $\rm J/\psi$ particles is observed.
 The acceptance and efficiency corrections are based on 
a detailed simulation of the detector response and the combined correction factor is about 
$10$. The main systematic error sources are: signal extraction ($\pm 8\%$), tracking and electron identification efficiency ($\pm 14.5\%$), minimum-bias cross section ($\pm 10\%$), and the unknown $\rm J/\psi$ polarization that affects the acceptance calculation by $^{+10}_{-25}\%$. The measured inclusive $\rm J/\psi$ 
production cross section is, thus:
$\d\sigma/\d y\,(|y|<0.88)=7.36\pm1.22(stat.)\pm1.32(syst.)^{+0.88}_{-1.84}(pol.)~\mu$b.

\begin{figure}[!t]
  \begin{center}
  \includegraphics[width=0.49\textwidth]{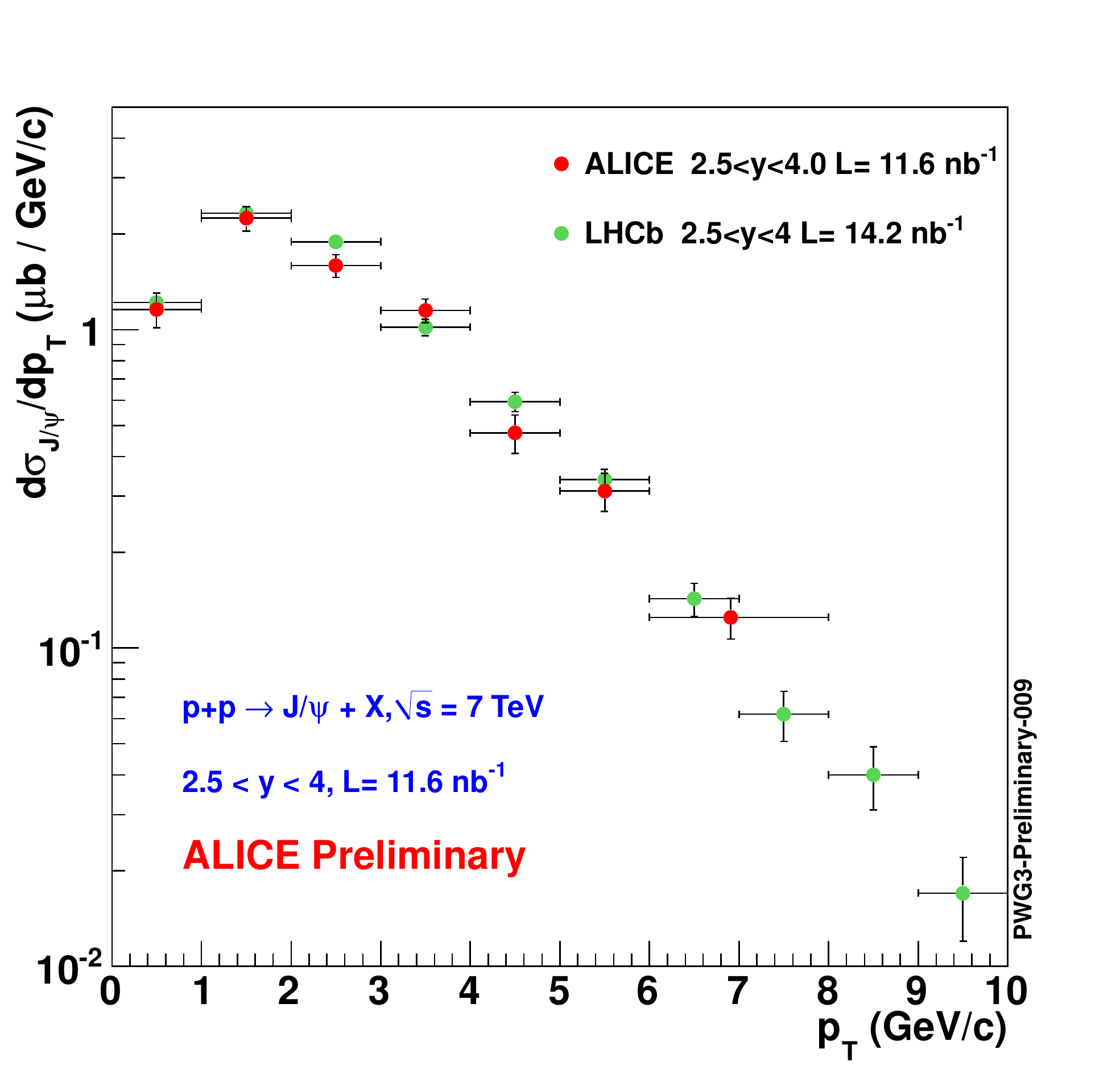}
  \includegraphics[width=0.49\textwidth]{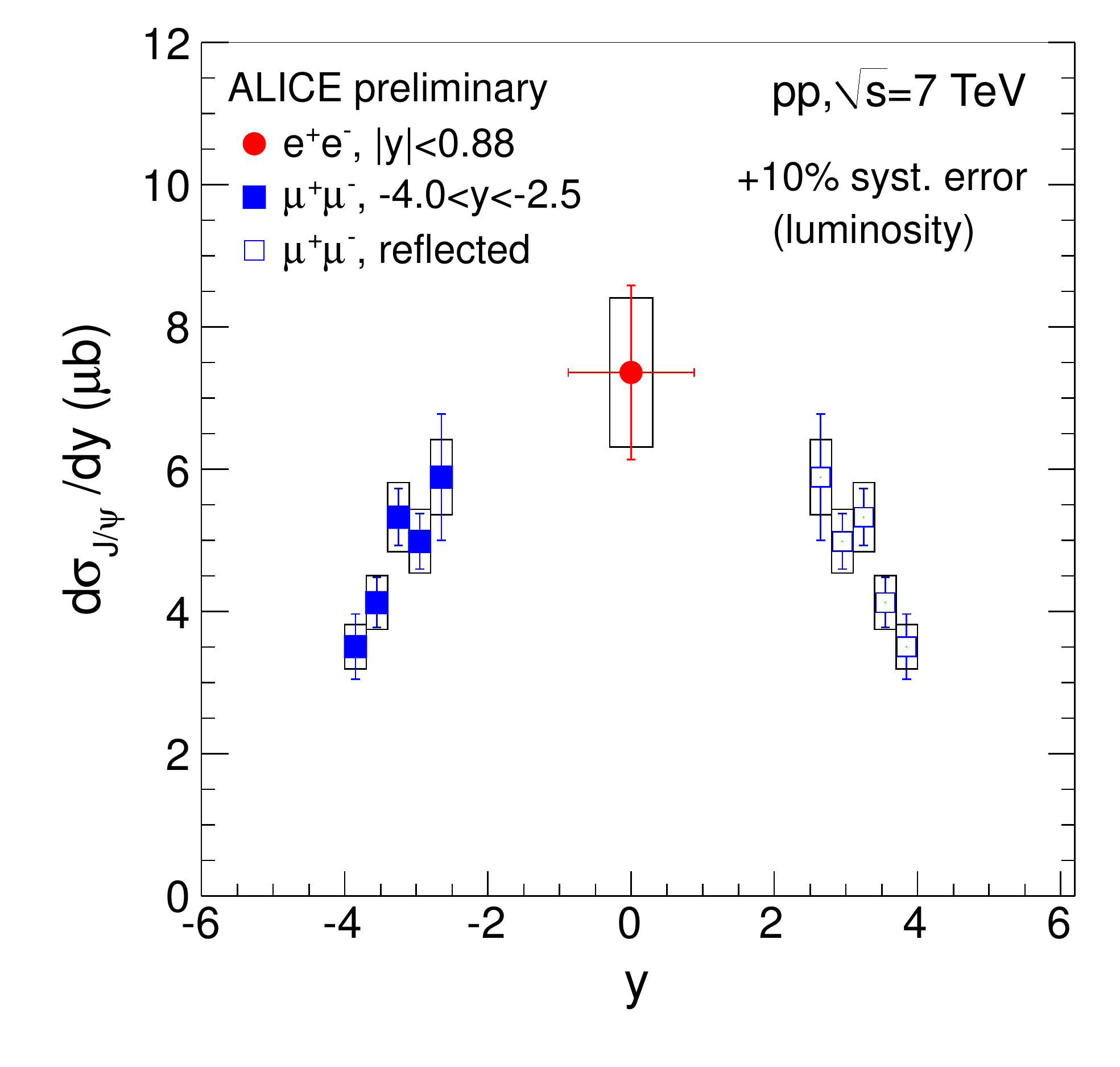}
  \caption{Inclusive $\rm J/\psi$ production cross section measured in pp collisions at $\sqrt{s}=7~\tev$, as a function of $\pt$ in the range $-4<y<-2.5$ using the di-muon decay channel (left) and as a function of rapidity in the range $\pt>0$ using also the di-electron decay channel (right). The inclusive  cross section measured by the LHCb experiment in the same rapidity range~\cite{lhcbICHEP} is reported for comparison in the left-hand panel.}
\label{fig:jpsi}
\end{center}
\end{figure}

The forward rapidity analysis, using di-muons reconstructed in the spectrometer, is based on 
a sample of $\rm 11.6~nb^{-1}$. The $\rm J/\psi$ signal is extracted by fitting
the di-muon invariant mass distribution with two Crystal Ball functions (for the $\rm J/\psi$ and the $\psi\prime$) and two exponentials (below and above the $\rm J/\psi$ mass), 
as described in Ref.~\cite{roberta}. The total $\rm J/\psi$ yield 
($\pt>0$ and $-4<y<-2.5$) is $1924\pm 77(stat.)$.
Also in this case, 
the acceptance and efficiency corrections are based on detailed detector simulation.
The main sources of systematic errors are: the unknown polarization ($^{+12}_{-21}\%$) and the minimum-bias cross section ($\pm 10\%$). 
The measured inclusive $\rm J/\psi$ 
production cross section is, thus:
$\sigma(-4<y<-2.5)=7.25\pm0.29(stat.)\pm0.98(syst.)^{+0.87}_{-1.50}(pol.)~\mu$b.
In order to measure the rapidity- and $\pt$-differential cross section, the invariant-mass
analysis is also performed
in seven bins in the transverse momentum range $0<\pt<8~\gev/c$ or in five bins in the rapidity range $-4<y<-2.5$. 
Figure~\ref{fig:jpsi} (left) shows $\d\sigma/\d\pt$ in $-4<y<-2.5$ (with statistical errors only),
compared to the corresponding measurement by the LHCb experiment~\cite{lhcbICHEP}.

The measurements in the di-muon channel, in five rapidity bins, and in the di-electron channel
(one point at central rapidity) allow to extract the $\d\sigma/\d y$ of inclusive $\rm J/\psi$
production for $\pt>0$ in the range $-4<y<0.88$, which is shown in the right-hand panel
of Fig.~\ref{fig:jpsi}. The next steps in these analyses, using the entire 2010 data sample, are: the measurement of $\d\sigma/\d\pt$ at central rapidity, and, at forward rapidity, 
the extension to higher $\pt$, the study of $\rm J/\psi$ production as a function of event multiplicity, the measurement of the $\psi\prime$ cross section, and of 
the $\rm J/\psi$ polarization. In addition, the fraction of $\rm J/\psi$ from B meson decays 
will be measured at central rapidity, exploiting the high-precision vertexing provided by the 
Inner Tracking System.

\section{Summary}

We have presented the first heavy-flavour production measurements performed by the 
ALICE experiment in proton--proton collisions at $\sqrt{s}=7~\tev$ at the LHC.
The production cross section of heavy-flavour decay muons at forward rapidity 
($-4<y<-2.5$) has been measured as a function of $\pt$ in the range 2--6.5~$\gev/c$.
Perturbative QCD calculations (FONLL) agree with our measurement within uncertainties.
The measurement of charm and beauty production at central rapidity is in progress, 
using fully-reconstructed decays of D mesons in hadronic final states and inclusive semi-electronic
decays of D and B mesons. These analyses are well advanced: the $\rm D^0$ and $\rm D^+$
$\pt$ distributions have been obtained and their ratio agrees with 
previous measurements; the inclusive electron $\pt$ distribution has been measured and it shows
a large excess, increasing with $\pt$, over a cocktail of non-heavy-flavour background sources.
We have also presented the first results on charmonium production, namely the rapidity- and
$\pt$-differential cross sections of inclusive $\rm J/\psi$. These measurements demonstrate the
points of strength of the ALICE design in this respect, i.e., the broad acceptance in rapidity 
and the sensitivity down to zero $\pt$ at all rapidities.




\end{document}